\documentclass{raa}

\usepackage{amssymb,amsmath}

\usepackage{graphicx,times}             %for PS/EPS graphics inclusion, new

\begin{document}

   \title{Accretion and jet power in active galactic nuclei}

   \volnopage{Vol.0 (200x) No.0, 000--000}      %%preserved for Editor. DOn't remove!
   \setcounter{page}{1}          %%starting page, preserved for Editor. DOn't remove!

   \author{L. Foschini
%      \inst{1}
   }

   \institute{INAF -- Osservatorio Astronomico di Brera, Via E. Bianchi 46, 23807, Merate (LC), Italy; {\it luigi.foschini@brera.inaf.it}\\
%% Please give the E-mail address of the author, to whom future correspondence and
%% offprint requests will be sent.
   }

   \date{Received~~2011 month day; accepted~~2011~~month day}

\abstract{The classical diagrams of radio loudness and jet power as a function of mass and accretion rate of the central spacetime singularity in active galactic nuclei are reanalyzed by including the data of the recently discovered powerful relativistic jets in Narrow-Line Seyfert 1 Galaxies. The results are studied in the light of the known theories on the relativistic jets, indicating that while the Blandford-Znajek mechanism is sufficient to explain the power radiated by BL Lac Objects, it fails to completely account the power from quasars and Narrow-Line Seyfert 1. This favors the scenario outlined by Cavaliere \& D'Elia of a composite jet, with a magnetospheric core plus a hydromagnetic component emerging as the accretion power increases and the disc becomes radiation-pressure dominated. A comparison with Galactic compact objects is also performed, finding some striking similarities, indicating that as the neutron stars are the low-mass jet systems analogue of black holes, the Narrow-Line Seyfert 1 Galaxies are the low-mass counterpart of the blazars.
}

%   \authorrunning{A.-Y. Zhou, E. Rodriguez \& B. J. Smith }            %author_head in even pages
%   \titlerunning{Photometry of $\delta$ Sct and Related Stars (I) }  % title_head in odd pages

\maketitle

\section{Introduction}
Despite their omnipresence in the Universe and thousands of written papers, relativistic jets are still poorly understood and there is not yet consensus on the mechanisms at work. 

In the discussion after the Blandford's seminal talk at the 1978 ``Pittsburgh conference on BL Lac Objects'', G. Burbidge raised one question about the possible importance of the host galaxy in the generation of relativistic jets in BL Lacs. Blandford answered that the host galaxy should not be a relevant detail, because the phenomena related to the jet occur within the central ten parsecs (Blandford \& Rees 1978). 

However, later observations seemed to support the idea that instead the host galaxy plays some role, with jets preferring ellipticals rather than spirals. By inverting the Blandford's answer, Laor (2000) asked how is it possible that the jet, which indeed is originated in the very inner part of a galaxy, could be related to the host. He suggested that one possible solution is that jets require large black hole masses ($\gtrsim 10^{9}M_{\odot}$), which in turn are hosted in ellipticals. On the other hand, AGN with no jets have masses of the central singularity $\lesssim 3\times 10^8M_{\odot}$. Sikora et al. (2007) confirmed some similar division, although with smoother boundaries, by finding that AGN with $M \gtrsim 10^8M_{\odot}$ have the radio loudness parameters 3 orders of magnitudes greater than those AGN with $M\lesssim 3\times 10^7M_{\odot}$.  

Sikora et al. (2007) reported also about some differences between the radio loudness and accretion rate of the central black hole in spiral- and elliptical-hosted AGN, where just a very few exceptions of spiral-hosted AGN can display high accretion and high radio loudness. Generally, the radio loudness is greater as the accretion rate is lower, somehow recalling the jets in Galactic binaries, which are linked to low/hard states (see Belloni 2010 for recent reviews).

Recently, Broderick \& Fender (2011) suggested a correction of the radio loudness due to the mass, which determines the vanishing of the radio-loud/quiet dichotomy and leaving only a general trend of a greater radio loudness as the accretion rate decreases.

Last, but not least, when speaking about jets, the black hole spin cannot be missing. It is like it has ``a finger in every pie'' (``come il prezzemolo'', in Italian saying), with its contribution mixed in every possible way. 

This scenario has been perturbed by the recent discovery of high-energy $\gamma$ rays from Narrow-Line Seyfert 1 Galaxies ($\gamma$-NLS1s, see Foschini 2011a for a recent review\footnote{See also the Proceedings of the workshop ``Narrow-Line Seyfert 1 Galaxies and Their Place in the Universe'' (Milano, Italy, April 4-6, 2011) to have a recent summary of the knowledge on NLS1s: \texttt{http://pos.sissa.it/cgi-bin/reader/conf.cgi?confid=126}}). These AGN have relatively small masses ($10^{6-8}M_{\odot}$), high accretion rate ($0.1-1L_{\rm Edd}$, see Fig.~8 in Foschini 2011a), are generally hosted by spiral galaxies, and could develop powerful relativistic jets, as luminous as those in blazars. Other differences with blazars refer to the full width half maximum (FWHM) of the broad permitted lines (in the case of H$\beta$, the value is $\gtrsim 2000$~km/s for blazars and $\lesssim 2000$~km/s for $\gamma$-NLS1s) and the radio morphology, which is very compact ($\lesssim 10$~pc, see e.g. Gu \& Chen 2010, except one known case, PKS~0558-504, Gliozzi et al. 2010) for $\gamma$-NLS1s, while blazars display extended structures, up to hundreds of parsecs. Therefore, even though the jets of $\gamma$-NLS1s are very similar to those in blazars, almost all the remaining part of the AGN and the host galaxy are different, meaning that $\gamma$-NLS1s are indeed a new class of $\gamma$-ray AGN(\footnote{Obviously, NLS1s are not new as AGN, since they were discovered by Osterbrock \& Pogge (1985) more than one quarter century ago. But NLS1s are new as $\gamma$-ray emitters. The difference seems subtle, but it is important. Just to cite Mark Twain: ``The difference between the right word and the nearly right word is the difference between lightning and lightning-bug''.}).

In the present work, I report on a comparative study of the main characteristics of the jets of blazars (BL Lac Objects and flat-spectrum radio quasars, FSRQs) and $\gamma$-NLS1s. I adopt a $\Lambda$CDM cosmology with $H_{0}=70$~km~s$^{-1}$~Mpc$^{-1}$ and $\Omega_{\Lambda}=0.73$ (Komatsu et al. 2011).

\begin{table}[!t]
\caption{Source list in alphabetical order per type (B: BL Lac Object; Q: FSRQ; S: $\gamma$-NLS1). The mass, disc luminosity, and jet power [erg~s$^{-1}$] are from Ghisellini et al. (2010) for the blazars and Abdo et al. (2009a) for the $\gamma$-NLS1. The radio flux density [Jy] are from the MOJAVE Project or from NED (indicated with *; in the case of PKS 1454$-$354 I adopted the value from Abdo et al. 2009b). The $\gamma$-NLS1 in italics are the new discoveries reported by Foschini (2011a). In this case, the jet power has been calculated with the correlation reported in Sect.~4. The last column indicates the radio-loudness parameter.}
\begin{center}
\scriptsize
\begin{tabular}{lccccccc}
\hline
\hline
Name & Type & $z$ & $\log M/M_{\odot}$ & $L_{\rm disc}/L_{\rm Edd}$ & $S_{\rm 15~GHz}$ & $\log P_{\rm jet}$ & RL\\
\hline
\hline
AO~0235$+$164 & B &  0.940 &   9.00 & $3.0\times 10^{-2}$  &  3.486 & 46.67 & 56784\\
BL~Lac        & B &  0.069 &   8.70 & $<4.0\times 10^{-4}$ &  4.122 & 44.97 & 26876\\
Mkn~421       & B &  0.031 &   8.70 & $3.5\times 10^{-6}$ &  0.327 & 43.71  & 47394\\
Mkn~501       & B &  0.034 &  8.84 & $5.7\times 10^{-6}$ &  0.848 & 43.48   & 63489\\
OJ~287        & B &  0.306 &   8.70 & $<3.0\times 10^{-3}$ &  4.587 & 45.15 & 97424\\
PKS~0332$-$403 & B &  1.445 &   9.70 & $8.0\times 10^{-2}$ &  1.781(*) & 47.12 & 7060\\
PKS~0537$-$441 & B &  0.894 &   9.30 & $4.0\times 10^{-2}$ &  10.667(*) & 47.22 & 57107\\
PKS~2155$-$304 & B &  0.116 &   8.90 & $<1.1\times 10^{-6}$ &  0.160(*) &  44.97 & 700894\\
S5~1803$+$784  & B &  0.680 &   8.70 & $2.5\times 10^{-2}$ &  2.709 & 46.82 & 46052\\
\hline
3C~273         & Q &  0.158 &   8.90 & $4.0\times 10^{-1}$ &  24.002 & 47.79 & 558\\
3C~279         & Q &  0.536 &  8.90 & $2.5\times 10^{-2}$ &  12.771 & 46.33 & 75558\\
3C~454.3       & Q &  0.859 &   9.00 & $2.0\times 10^{-1}$ &  15.864 & 47.89 & 30610\\
B2~1520$+$31   & Q &  1.487 &   9.40 & $1.5\times 10^{-2}$ &  0.402 &  46.62 & 18441\\
B2~1846$+$32A  & Q &  0.798 &   8.70 & $1.3\times 10^{-1}$ &  0.496 &  46.61 & 2434\\
B3~0650$+$453  & Q &  0.933 &   8.48 & $1.0\times 10^{-1}$ &  0.332 &  46.90 & 5302\\
B3~0917$+$449  & Q &  2.190 &  9.78 & $2.0\times 10^{-1}$ &  2.100 &  47.61 & 9427\\
B3~1633$+$382  & Q &  1.813 &   9.70 & $1.0\times 10^{-1}$ &  3.312 &  47.12 & 20254\\
PKS~0227$-$369 & Q &  2.115 &   9.30 & $1.0\times 10^{-1}$ &  0.287(*) & 47.38 & 6955\\
PKS~0347$-$211 & Q &  2.994 &   9.70 & $1.0\times 10^{-1}$ &  0.474 & 47.05 & 13560\\
PKS~0454$-$234 & Q &  1.003 &   9.40 & $5.0\times 10^{-2}$ &  1.820 & 46.55 & 8452\\
PKS~1454$-$354 & Q &  1.424 &   9.30 & $1.5\times 10^{-1}$ &  1.230(*) & 47.60 & 6236\\
PKS~1502$+$106 & Q &  1.838 &   9.48 & $1.3\times 10^{-1}$ &  1.641 & 47.07 & 13397\\
PKS~2023$-$07  & Q &  1.388 &   9.48 & $5.0\times 10^{-2}$ &  1.005 & 46.94 & 9480\\
PKS~2144$+$092 & Q &  1.113 &   9.00 & $1.0\times 10^{-1}$ &  0.845 & 47.01 & 6490\\
PKS~2201$+$171 & Q &  1.076 &   9.30 & $4.0\times 10^{-2}$ &  1.088 & 46.78 & 9532\\
PKS~2204$-$54  & Q &  1.215 &   9.00 & $1.8\times 10^{-1}$ &  1.277(*) & 47.11 & 6926\\
PKS~2227$-$08  & Q &  1.559 &   9.18 & $1.1\times 10^{-1}$ &  5.158 & 47.31 & 61576\\
PKS~B0208$-$512 & Q & 0.999 &   8.84 & $1.4\times 10^{-1}$ &  2.893(*) & 46.49 & 16954\\
PKS~B1127$-$145 & Q & 1.184 &   9.48 & $2.5\times 10^{-1}$ &  2.558 & 47.32 & 3101\\
PKS~B1508$-$055 & Q & 1.185 &   9.30 & $2.0\times 10^{-1}$ &  0.769 & 46.63 & 1752\\
PKS~B1510$-$089 & Q & 0.360 &   8.84 & $4.0\times 10^{-2}$ &  2.401 & 46.79 & 3960\\
PKS~B1908$-$201 & Q & 1.119 &   9.00 & $2.0\times 10^{-1}$ &  6.727 & 46.92 & 26215\\
PMN~J2345$-$1555 & Q & 0.621 &  8.60 & $6.0\times 10^{-2}$ &  0.635(*) & 45.97 & 4485\\
S3~2141$+$17 & Q & 0.211 & 8.60 &  $1.2\times 10^{-1}$ & 0.942 & 45.28 & 273\\
S4~0133$+$47 & Q & 0.859 & 9.00 &  $1.0\times 10^{-1}$ & 3.536 & 47.11 & 13646\\
S4~0954$+$55 & Q & 0.895 & 9.00 &  $2.0\times 10^{-2}$ & 0.210 & 45.46 & 4517\\
S4~1030$+$61 & Q & 1.400 & 9.48 & $4.0\times 10^{-2}$ & 0.400 & 46.69 & 4832\\
S4~1849$+$67 & Q & 0.657 & 8.78 & $5.0\times 10^{-2}$ &  2.700 & 46.41 & 17548\\
SBS~0820$+$560 & Q & 1.418 & 9.18 & $1.5\times 10^{-1}$ & 1.682(*) & 46.91 & 11236\\
\hline
1H~0323$+$342 & S & 0.061 & 7.00 &  $9.0\times 10^{-1}$ &  0.353 & 44.36 & 40\\ 
\emph{FBQS~J1102$+$2239} & S & 0.453 & 7.62 & $4.0\times 10^{-1}$ & 0.003 & 44.57 & 13\\
PKS~1502$+$036 & S & 0.409 & 7.30 & $8.0\times 10^{-1}$ & 0.496 & 46.21 & 1926\\
PKS~2004$-$447 & S & 0.240 & 6.70 & $2.0\times 10^{-1}$ & 0.227(*) & 44.16 & 4198\\
PMN~J0948$+$0022 & S & 0.585 & 8.18 & $4.0\times 10^{-1}$ & 0.473 & 46.72 & 1153\\
\emph{SBS~0846$+$513} & S & 0.584 & 7.56 & $4.7\times 10^{-1}$ & 0.225 & 46.34 & 1937\\
\emph{SDSS~J1246$+$0238} & S & 0.363 & 7.34 & $7.6\times 10^{-1}$ & 0.036 & 45.30 & 102\\
\hline
\end{tabular}
\end{center}
\label{tab:sourcelist}
\normalsize
\end{table}

\section{Sample selection}
I have collected the data of 30 FSRQs and 9 BL Lac Objects from Ghisellini et al. (2010) and I added the 4 $\gamma$-NLS1s from Abdo et al. (2009a). The total jet power reported in those papers has been calculated by means of the spectral energy distribution (SED) modeling, according to Ghisellini \& Tavecchio (2009). The same model calculates the masses and the accretion rates by fitting the optical/ultraviolet emission to a standard Shakura-Sunyaev accretion disc and the results have been cross-checked with the measurements made with other independent methods available in literature (mostly by the classical reverberation mapping technique). 

Some BL Lacs have just an upper limit for the disc luminosity. Recently, weak Ly$\alpha$ emission lines (equivalent width $<<1$~\AA) have been observed in Mkn 501 and Mkn 421 (Stocke et al. 2011). The disc luminosity [in Eddington units] calculated from these lines, by assuming that $L_{\rm disc}\sim 10L_{\rm Ly\alpha}$, is $5.7\times 10^{-6}$ and $3.5\times 10^{-6}$ for Mkn 501 and Mkn 421, respectively (see Foschini 2011b). The non-detection of any line in PKS~2155$-$304 poses an upper limit tighter than the upper limit in Ghisellini et al. (2010). Therefore, in these three cases (Mkn 421, Mkn 501, PKS~2155$-$304), I adopt the values of the accretion from Stocke et al. (2011).

Radio data have been taken from the \emph{Monitoring Of Jets in Active galactic nuclei with VLBA Experiments} (MOJAVE) at 15~GHz (Lister et al. 2009). When the source is not in the MOJAVE list, then data from NED(\footnote{\texttt{http://ned.ipac.caltech.edu/}}) have been used. In case of multiple measurements, then the average was calculated. When possible, the measurements performed in the period covered by \emph{Fermi}/LAT observations were considered. If no 15~GHz measurements were available, then data at the nearest frequency have been selected and converted to 15~GHz by adopting a flat radio spectral index ($\alpha_{\rm r}=0$). The radio loudness parameter $RL$ has been calculated by using the radio flux at 15~GHz and the ultraviolet flux at 200~nm, calculated from the accretion luminosity. Basically, $RL$ is now a better indicator of the jet dominance over the accretion. 

To increase the statistics of $\gamma$-NLS1s, I have included in the sample also the three newly discovered sources reported by Foschini (2011a). In this case, no modeling is available and the jet power has been calculated by using the ``calibration'' of the Ghisellini \& Tavecchio (2009) model (see Sect. 4). 

The source list is shown in Table~\ref{tab:sourcelist}.

\begin{figure}[!t]
\begin{center}
\includegraphics[angle=270,scale=0.4]{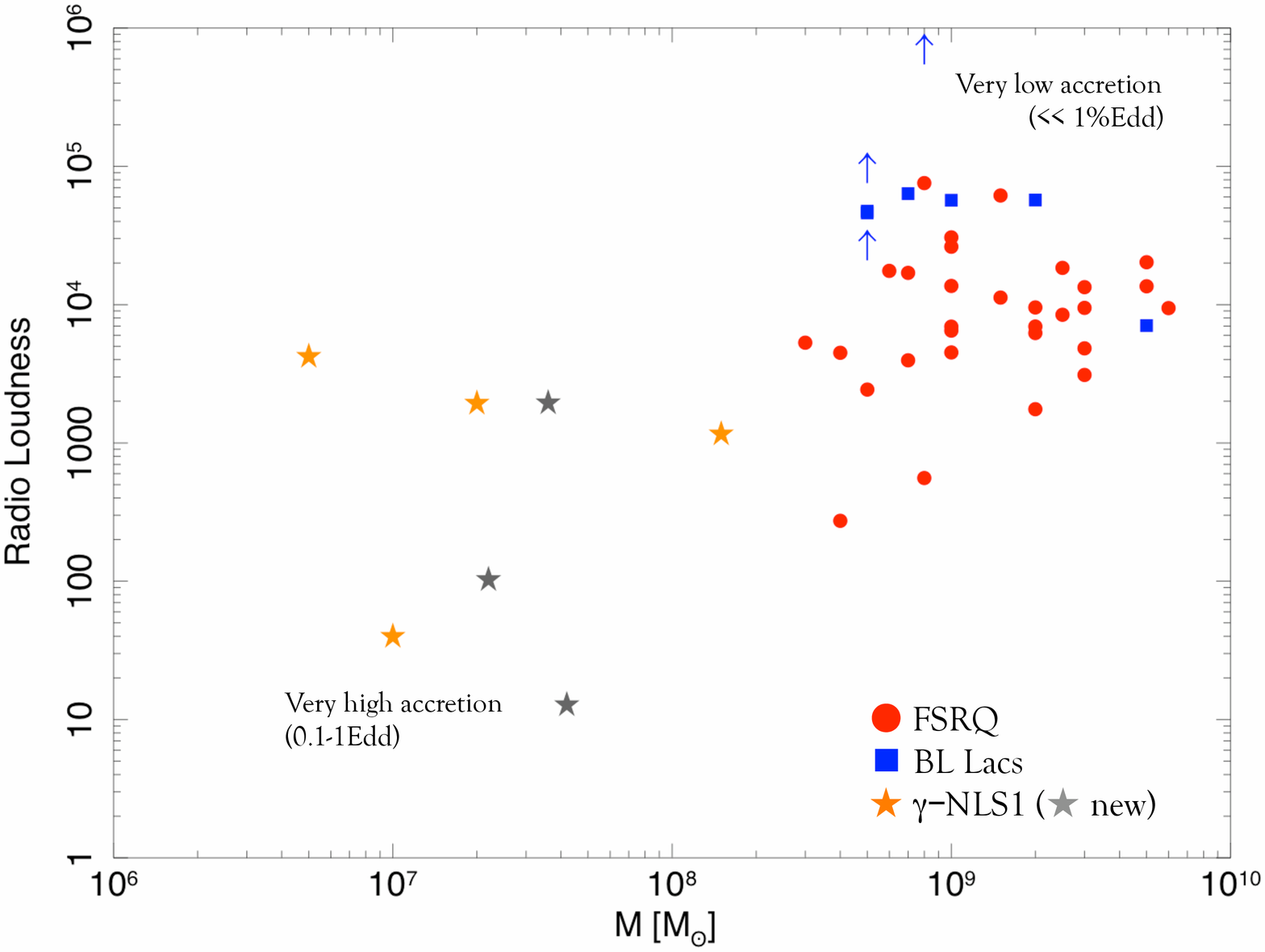}\\
\includegraphics[angle=270,scale=0.4]{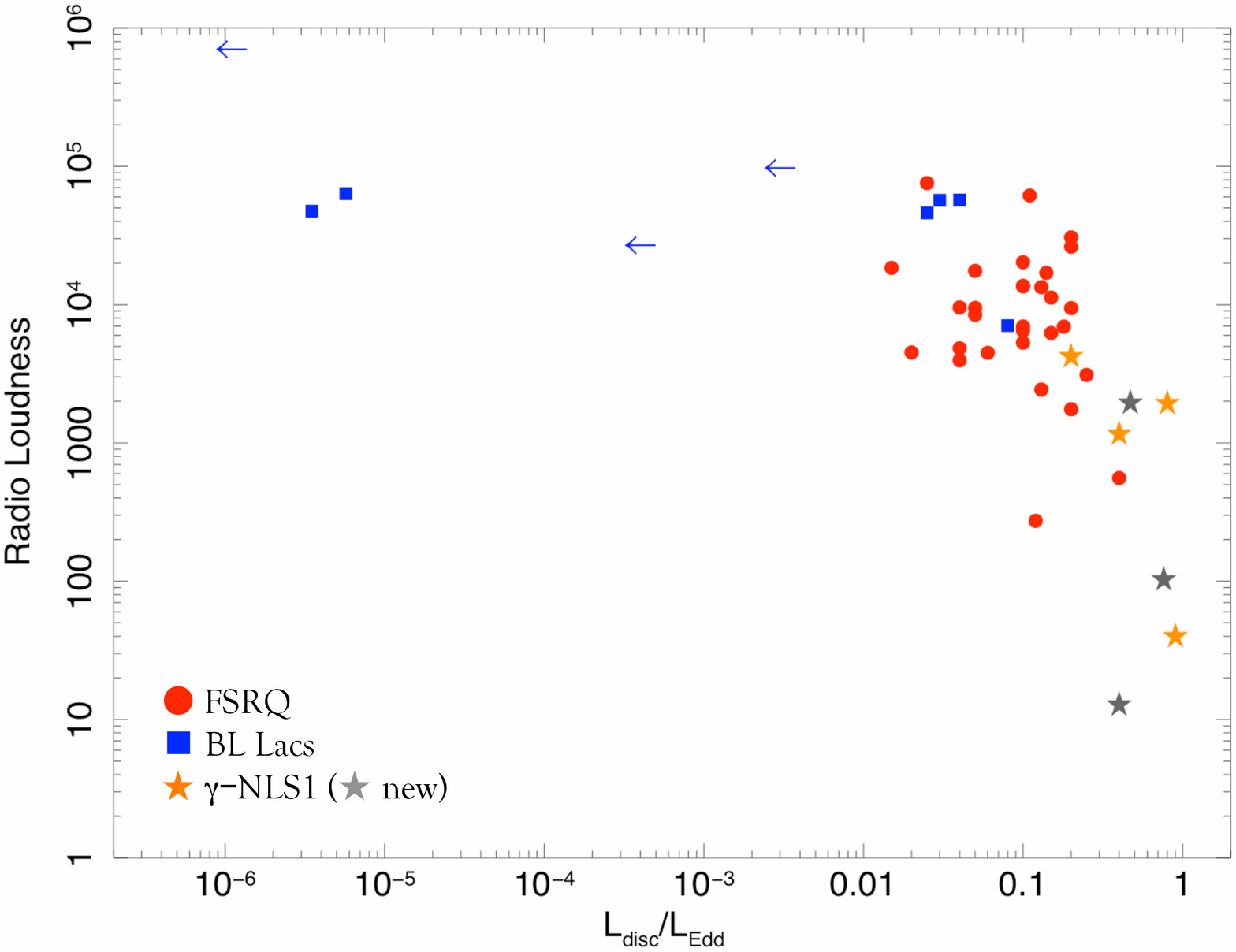} 
\caption{Radio loudness vs mass of the central black hole (\emph{top panel}) and disc luminosity in Eddington units (\emph{bottom panel}).}
\label{fig:radioloudness}
\end{center}
\end{figure}

\section{Who cares about radio loudness?}
Fig.~\ref{fig:radioloudness} display the radio loudness vs the mass of the central black hole and its accretion luminosity in Eddington units, which in turn is proportional to the accretion rate. Fig.~\ref{fig:radioloudness} (\emph{top panel}) has to be compared with Fig.~4 of Sikora et al. (2007) and Fig.~1 of Woo \& Urry (2002). The former reported a dependence of $RL$ with the mass (high $RL$ requires high masses), while the latter did not find this dependence. The sample studied in the present work is composed of AGN with jets, so it is not exactly comparable with the larger samples of Sikora et al. (2007) and Woo \& Urry (2002), but some interesting features are already visible. 

I note a central region, broadly consistent with the results of Woo \& Urry (2002). The $\gamma$-NLS1s are populating the region with high $RL$ and low masses. The deviations from this central zone are quite localized in two regions: one where are objects with high masses, low accretion rates, but with high $RL$ (featureless BL Lacs); the other refers to objects with low masses, high accretion rates and low $RL$. This seems to suggest that the changes in the radio loudness are more linked to the accretion rate, rather than directly to the mass (see, however, Sect.~4). This is indeed what is shown in Fig.~\ref{fig:radioloudness} (\emph{bottom panel}), where is shown a trend of decreasing $RL$ with increasing accretion rate, which is in agreement with the results shown in Fig.~3 of Sikora et al. (2007) and Fig.~1 of Broderick \& Fender (2011). 

It is worth noting that the sources with high and low $RL$ have low power jets. So, high $RL$ does not mean a powerful jet, but rather a low accretion. One could ask oneself if the radio loudness is still a meaningful parameter. The answer could be yes, if one wants just to know if an AGN has a jet or not. No, if one wants to perform a deeper study of relativistic jets.

\begin{figure}[!t]
\begin{center}
\includegraphics[angle=270,scale=0.4]{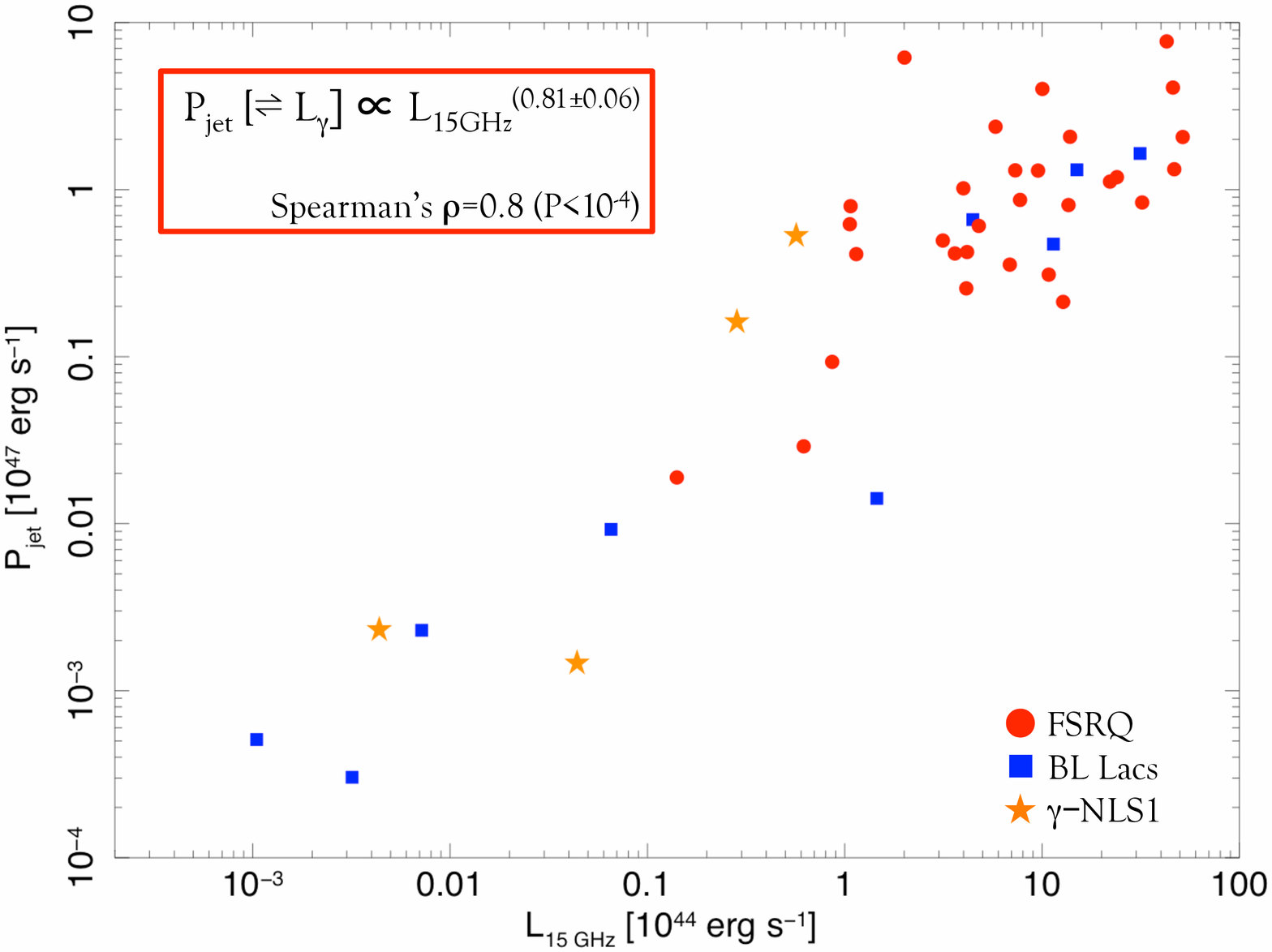} 
\caption{Jet power vs radio power at 15~GHz.}
\label{fig3}
\end{center}
\end{figure}

\section{Jet power}
Before going ahead, it is worth performing some kind of ``calibration'' of the total jet power calculated according to the model by Ghisellini \& Tavecchio (2009) by adding the radiative, magnetic, and particles contributions. It is known that the luminosity of a relativistic jet is correlated with its radio core emission according to (Blandford \& K\"onigl 1979, K\"ording et al. 2006):

\begin{equation}
P_{\rm jet} \propto L_{\rm radio,core}^{12/17}
\label{theorjet}
\end{equation}

\noindent and, therefore, this relationship can be used as a way to ``calibrate'' the Ghisellini \& Tavecchio's model. In this work, I used the radio data at 15~GHz, mostly from the MOJAVE Project, which in turn is based on high-resolution VLBA observations that allow to have the best measurement of the core available to date. Obviously, in this case the three new $\gamma$-NLS1s (sources in italics in Table~\ref{tab:sourcelist}) have not been considered, since no modeling is available yet. Instead, the relationship obtained and shown below has been used to calculate the jet power of these three sources from the radio measurements. 

The results are displayed in Fig.~\ref{fig3}, where the correlation is already visible by eye. The search for correlation has been performed by using the \texttt{ASURV Rev.~1.2} software package (Lavalley et al. 1992), which makes use of the algorithms by Feigelson \& Nelson (1985) and Isobe et al. (1986). The correlation is confirmed by the Spearman's $\rho=0.8$ ($P_{\rm chance}<10^{-4}$) and a high Z value of the Kendall's test ($Z=5.6$, $P_{\rm chance}<10^{-4}$). The two powers are linked by the following equation:

\begin{equation}
\log P_{\rm jet} = (11\pm3) + (0.81\pm 0.06)\log L_{\rm 15~GHz}
\label{calibrationGG}
\end{equation}

The index $0.81$ is not exactly consistent with the theoretical value of $12/17(\sim 0.71)$ in the Eq.~(\ref{theorjet}), but the difference is not very significant by taking into account the error. In addition, it is worth noting that the present sample is basically built on a limited sample. When more data will be added, the correlation is likely to improve. It is possible to have an idea of the robustness of the result, by cross-cheking with similar results obtained by other authors with more complete samples. Since the jet power is directly linked to the power emitted at high-energy $\gamma$ rays, it is possible to compare the Eq.~(\ref{calibrationGG}) with other correlation between radio and $\gamma$ rays. For example, Bloom (2008) found: 

\begin{equation}
L_{\rm 400~MeV}\propto L_{\rm 8.4~GHz}^{0.77\pm0.03}
\end{equation}

\noindent while Ghirlanda et al. (2011) found a steeper index, but by using the integrated luminosity at $E>100$~MeV:

\begin{equation}
L_{\rm >100~MeV}\propto L_{\rm 20~GHz}^{1.07\pm0.05}
\end{equation}

Therefore, I conclude that the ``calibration'' of the Ghisellini \& Tavecchio's model (Eq.~\ref{calibrationGG}) is reliable, despite of the above mentioned caveats. 

\begin{figure}[!ht]
\begin{center}
\includegraphics[angle=270,scale=0.38]{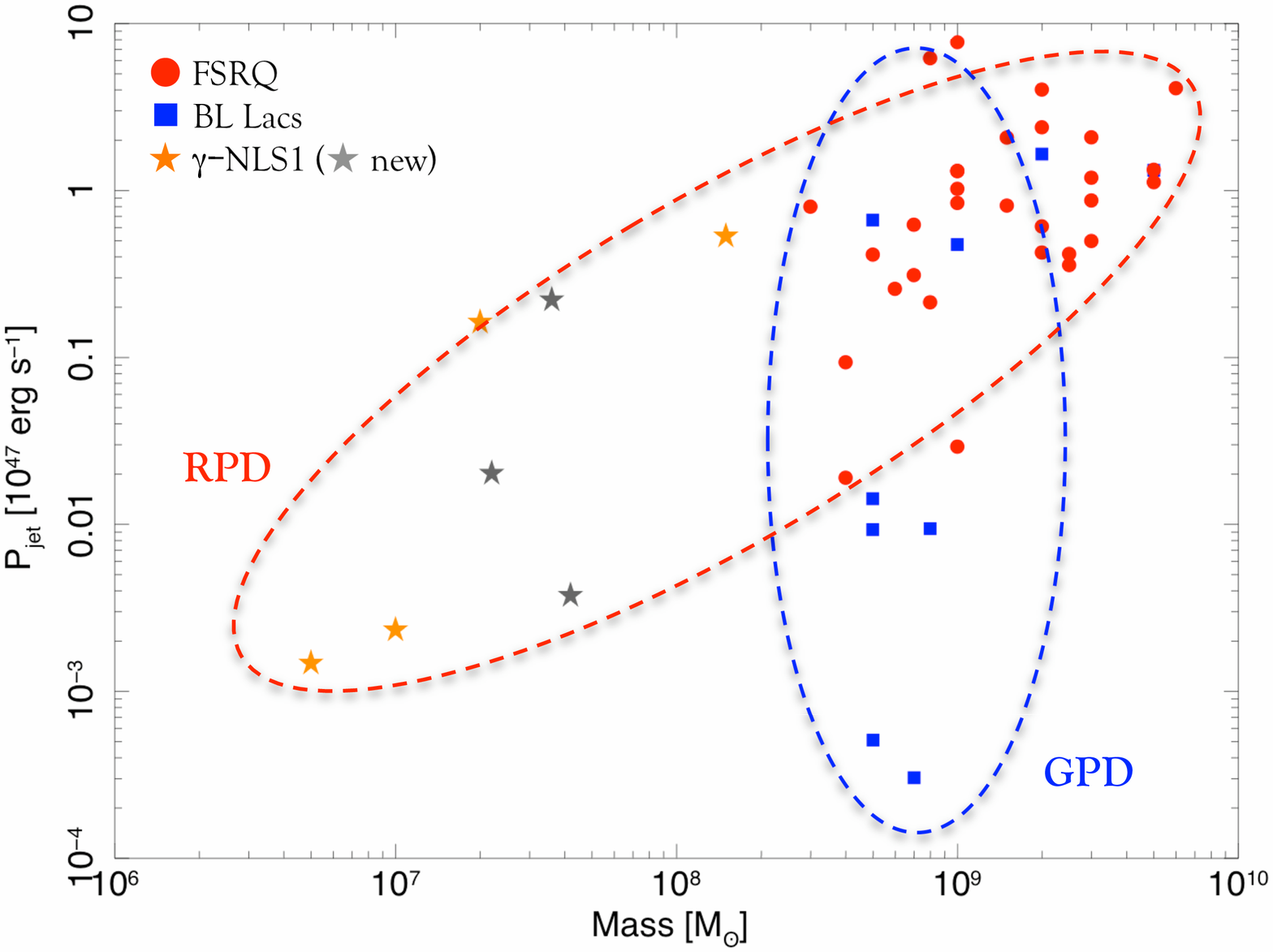}\\
\includegraphics[angle=270,scale=0.38]{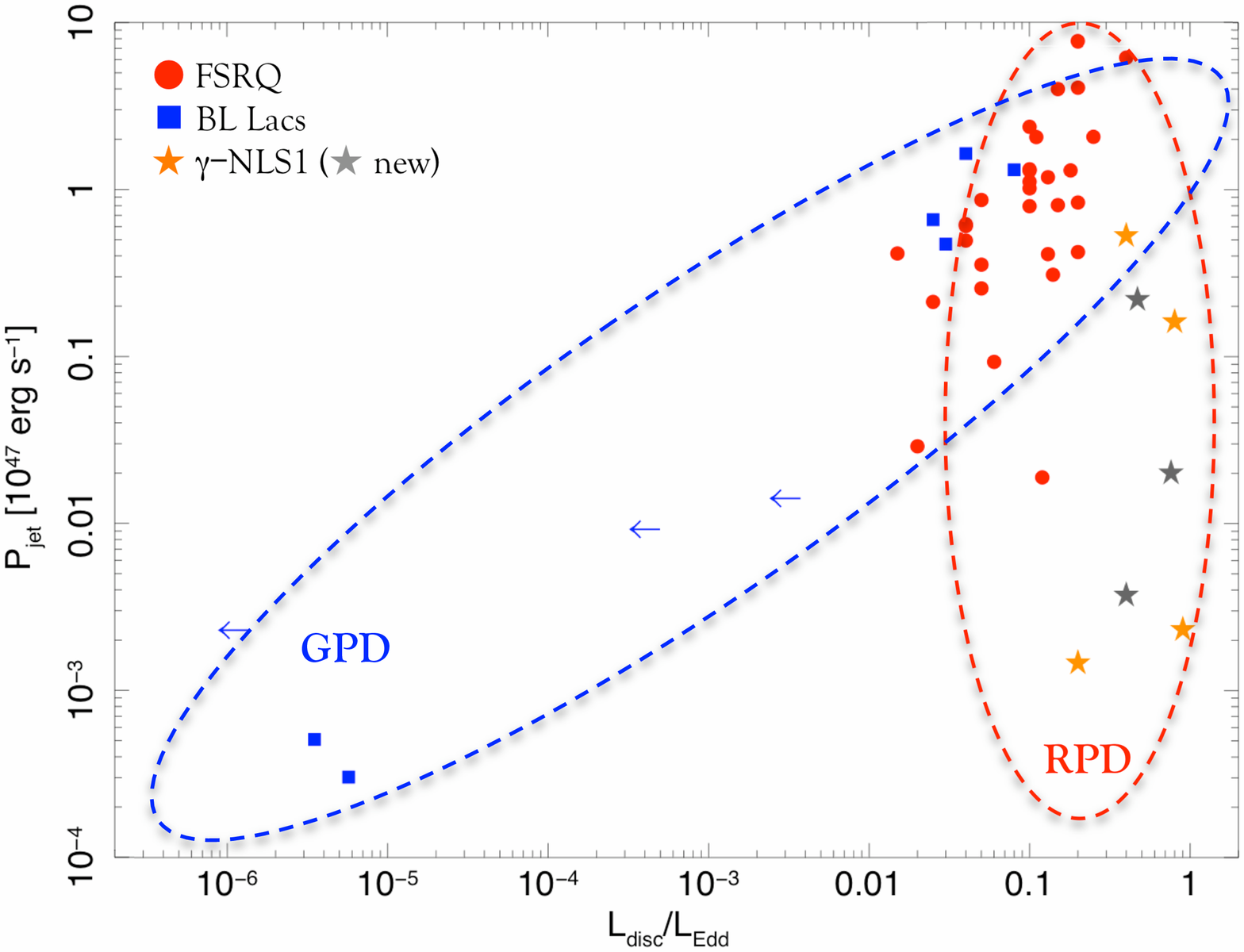} 
\caption{Jet power vs mass of the central black hole (\emph{top panel}) and the accretion luminosity in Eddington units (\emph{bottom panel}). See the text for more details.}
\label{fig:jetpower}
\end{center}
\end{figure}

It is then possible to check if there is any dependence of the jet power on the mass. Fig.~\ref{fig:jetpower} (\emph{top panel}) displays the corresponding plot. The presence of $\gamma$-NLS1s now fills the part of low masses. The dashed lines are just to remind the branches that are much more evident (but inverted!) in the graphic of the jet power vs the accretion luminosity in Eddington units, as displayed in Fig.~\ref{fig:jetpower} (\emph{bottom panel}). FSRQs and BL Lac Objects are placed along a line from low power/low accretion to high power/high accretion. This is the well-known ``blazar main sequence'' (see the blue dashed ellipse), where FSRQ have a strong disc and evolve to poorly accreting BL Lac Objects (Cavaliere \& D'Elia 2002, B\"ottcher \& Dermer 2002, Maraschi \& Tavecchio 2003). It can be read also as the ``blazar cooling sequence'' as revised by Ghisellini \& Tavecchio (2008), where the jet power is a function of the accretion rate. In this case, powerful discs determine a rich environment where electrons can cool efficiently (high power blazars), while on the opposite side there are BL Lacs with weak discs and hence a photon-starved environment, which in turn imply a greater difficulty to cool relativistic electrons. However, I note immediately than $\gamma$-NLS1s develop a separate branch (red dashed ellipse), where a low power jet is associated with a highly accreting disc. 

Also in this case, it is possible to compensate the small and non-homogeneous sample with the comparison with another larger sample. In this case, one can compare Fig.~\ref{fig:jetpower} (\emph{bottom panel}) of the present work with Fig.~5 of Ghisellini et al. (2011). The blazar part is confirmed, with the branch at low accretion where $L_{\rm disc}\propto \dot{m}^2$ and the high-accretion zone with $L_{\rm disc}\propto \dot{m}$, while the present work adds the $\gamma$-NLS1s branch. 

Fig.~\ref{fig:jetpower} shows also two different regimes: one where the jet power depends on the accretion and another where it scales with the mass. BL Lac Objects are in the accretion-dependent regime, while FSRQs and $\gamma$-NLS1s are in the mass-dependent regime. Indeed, it is easy to recognize that the difference of the jet power between FSRQs and $\gamma$-NLS1s can be explained by the difference of mass between the two classes of AGN. These two regimes reminds the known theories on jets. The labels RPD and GPD in Fig.~\ref{fig:jetpower} are the acronym of Radiation-Pressure Dominated and Gas-Pressure Dominated regimes and their meaning can be understood in the next Section.

\section{Magnetospheric, Hydromagnetic, Hybrid: the flavors of jets}
One advantage of the SED modeling is that it is possible to separate the different components of the power emitted by the jet: kinetic (particles), magnetic field, and radiation. Therefore, it is easier to compare with the theories. 

Basically, the known mechanisms of relativistic jets can be broadly divided into three classes:

\begin{enumerate}
\item \emph{Magnetospheric Jet:} the jet extracts rotational energy from the black hole and the accretion disc, through the slip between the magnetic fields at the hole and anchored at the accretion disc caused by the frame-dragging. Therefore, a black hole is needed to provide the Lense-Thirring effect, while the disc provides mainly the electric charges to be accelerated. The reference paper is Blandford \& Znajek (BZ, 1977), but also Macdonald \& Thorne (1982) represents an interesting alternative explanation of the same mechanism by adopting the membrane paradigm (Thorne et al. 1986). It can be considered the analogous for black holes of the pulsar magnetosphere by Goldreich \& Julian (1969). Some ``precursors'' of the BZ theory are Penrose (1969), Ruffini \& Wilson (1976), Lovelace (1976), who developed the analogy with an electric machine, and Blandford (1976), who elaborated an embryo version of the BZ effect in a flat spacetime. The BZ power depends on the mass of the black hole (which in turn affects the frame-dragging amplitude), the spin (magnetic fields slip), and the hole magnetic field. 

\item \emph{Hydromagnetic Jet:} it is a centrifugally-driven jet and it extracts the rotational energy of the accretion disc. There is no need of a black hole; only an accretion disc is necessary. The reference work is Blandford \& Payne (BP, 1982), with some precursors also in this case: Piddington (1970), Sturrock \& Barnes (1972), Ozernoy \& Usov (1973).  The power extracted in this way is proportional to the disc magnetic field, the size of the disc and its angular speed.

\item \emph{Hybrid models (``hydromagnetospheric''):} basically these models are a mixture of the two above cases. Interesting models are Phinney (1983) and Meier (1999), which in turn is an evolution of Punsly \& Coroniti (1990). A hybrid model has been recently adopted by Garofalo et al. (2010) to speculate on the observed differences in the AGN with relativistic jets.
\end{enumerate}

In the BZ theory, the magnetic field of the disc is pushed toward the event horizon by the Maxwell pressure. The standard disc magnetic field depends on the accretion rate $\dot{m}$ and it is possible to find two regimes (Ghosh \& Abramowicz 1997; Moderski \& Sikora 1996). One refers to strong accretion disc, dominated by the radiation pressure (RPD, radiation pressure dominated). The jet luminosity that can be extracted through the BZ mechanism saturates to the value (Ghosh \& Abramowicz 1997):

\begin{equation}
L_{\rm BZ,RPD} = 2\times 10^{44} M_8 (J/J_{\rm max})^2
\label{eq:bzrpd}
\end{equation}

\noindent where $M_{8}$ is the black hole mass in units of $10^{8}M_{\odot}$ and $J$ is the angular momentum of the hole. 

The second case is complementary to the first one and refers to a gas pressure dominated (GPD) disc (i.e. low accretion, Ghosh \& Abramowicz 1997):

\begin{equation}
L_{\rm BZ,GPD} = 8\times 10^{42} M_{8}^{11/10} \dot{m}_{-4}^{4/5} (J/J_{\rm max})^2
\label{eq:bzgpd}
\end{equation}

\noindent where $\dot{m}_{-4}$ is the accretion rate in units of $10^{-4}$. The dividing line of the two regimes is at $\dot{m}\sim 5.6\times 10^{-3}$, which corresponds to $L_{\rm disc}/L_{\rm Edd}\sim 5.6\times 10^{-4}$, by adopting the usual value for the efficiency equal to $\eta = 0.1$ (\footnote{The efficiency is generally dependent on the accretion luminosity in presence of advection, but the basic conceptual result does not change.}). From Fig.~\ref{fig:jetpower}, it is easy to recognize that BL Lacs are basically in the GPD regime (hence the jet power depends on the accretion, according to Eq.~\ref{eq:bzgpd}), while FSRQs and $\gamma$-NLS1s are in RPD (jet power scales with the mass, according to Eq.~\ref{eq:bzrpd}).

\begin{figure}[!t]
\begin{center}
\includegraphics[angle=270,scale=0.4]{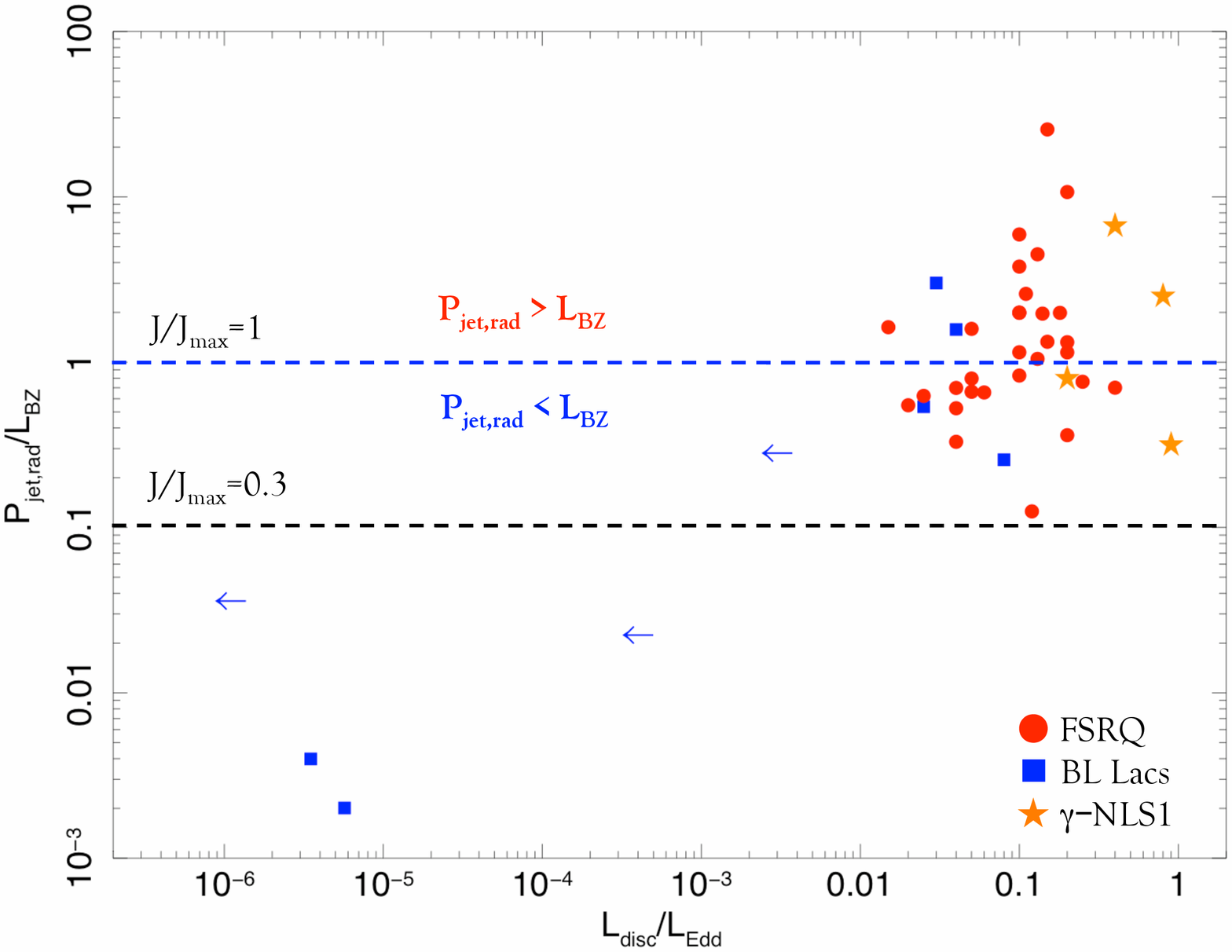} 
\caption{Ratio between the observed jet radiative power and the calculated luminosity according to the Blandford-Znajek theory. See the text for more details.}
\label{fig:BZratio}
\end{center}
\end{figure}

By taking into account the above caveats, it is therefore possible to calculate the jet luminosity with the Blandford-Znajek theory and compare with the observed radiative power dissipated calculated through the SED modeling(\footnote{In this case, I do not use the data of the three new $\gamma$-NLS1s because there is no SED model available yet.}). The results are shown in Fig.~\ref{fig:BZratio}. The BZ luminosity has been calculated according to Eqs.~(\ref{eq:bzrpd}) and (\ref{eq:bzgpd}), with the $\eta \propto \sqrt{L_{\rm disc}/L_{\rm Edd}}$ and $J/J_{\rm max}=1$ (see Foschini 2011c for some notes on the efficiency). It is therefore the maximum value of luminosity that can be produced by the Blandford-Znajek mechanism. As the luminosity of the disc decreases, the BZ mechanism is more than sufficient to produce the observed jet luminosity. The fact that $P_{\rm jet,rad}<L_{\rm BZ}$ can be explained by taking into account different values of $J/J_{\rm max}<1$ in Eqs.~(\ref{eq:bzrpd}) and (\ref{eq:bzgpd}). In addition, Eqs.~(\ref{eq:bzrpd}) and (\ref{eq:bzgpd}) of Ghosh \& Abramowicz (1997) have been elaborated by assuming a constant slip factor(\footnote{$s=\omega_{\rm F}(\omega_{\rm H}-\omega_{\rm F})/\omega_{\rm H}^2$, where $\omega_{\rm H}$ is the angular speed of the black hole and $\omega_{\rm F}$ is that of the magnetic field coming from the disc.}) $s$ equal to the maximum possible ($s=0.25$). This is the reference value adopted by most authors, but it refers to a radial magnetic field. As noted by Blandford \& Znajek (1977), a parabolic field has a slightly lower efficiency (75\% of the radial field). 

Anyway, the important information is that BL Lacs with low accretion could be essentially powered by the Blandford-Znajek mechanism or at least there is no need to invoke alternative or additional mechanisms. This is in agreement with the fact that these are objects at the end of their evolution, as already outlined by Cavaliere \& D'Elia (2002) and Maraschi \& Tavecchio (2003). The disc is weaker and weaker, but it is sufficient to provide some electric charges to be accelerated. The magnetic field of the hole is quite strong, since it is the result of the field that the disc has pushed toward the event horizon during its lifetime. 

As the disc luminosity increases, also the jet power increases and exceeds the BZ luminosity. By taking into account also the presence of the spin factor, it results that the observed luminosity overcomes the calculated one by $2-3$ orders of magnitudes, which it cannot be explained with the source variability or the errors in the measurements of parameters. Since this occurs at high disc luminosities, it is reasonable to expect the possibility of a contribution to the jet luminosity from hydromagnetic winds from the disc, thus creating some hybrid mechanism. The MHD luminosity according to the hybrid model by Meier (1999) could be more than three orders of magnitudes greater than that from a simple magnetospheric jet. 

Moreover, again the evolutionary path could play some role: both FSRQs and $\gamma$-NLS1s are young sources, with highly accreting discs, but although this is the right condition to trigger hydromagnetic winds, it could be also noted that the disc could have had not sufficient time yet to push a strong magnetic field to the hole and therefore the magnetospheric contribution to the jet is small. 

The present data do not allow to distinguish the contribution from the different mechanisms, but can confirm the broad scenario already outlined by Cavaliere \& D'Elia (2002), where the Blandford-Znajek is the backbone of the jet and becomes increasingly hybrid as the accretion increases. The two regimes, GPD and RPD (Ghosh \& Abramowicz 1997), are still valid, but with the warning that the jet mechanism becomes hybrid in the RPD regime. The strong disc, if on one side contributes to saturate the accretion regime of the BZ mechanism, on the other hand it enhance the jet power with the rise of hydromagnetic winds or the hybridization mechanisms. Being much more efficient (e.g. Meier 1999), but still dependent on the mass, it can explain why it is possible to find powerful jets even in small AGN, like $\gamma$-NLS1s. It is worth noting the discovery of ultra fast outflows with $v\sim (0.04-0.15) c$ in some highly accreting radio galaxies reported by Tombesi et al. (2010). Although these winds are moving at relativistic speeds, their discovery indicates that some hydromagnetic windy activity is present also in AGN with relativistic jets, thus enforcing the idea of a hybrid mechanism.

\section{Similarities with Galactic binaries jets} 
As known, relativistic jets are rather ubiquitous structures in the Universe and therefore it is necessary to find the basic laws driving how to scale the jet power with the mass, down to stellar mass compact objects. Recently, Coriat et al. (2011) made a detailed study on Galactic binaries of the correlation between the radio emission at 8.4~GHz, which stands for the jet power, and the X-ray luminosity ($3-9$~keV), which in turn samples the disc emission in the Galactic binaries (see Fig.~5 in Coriat et al. 2011). They identified two main branches: one is the ``inefficient'' branch, characterized by $L_{\rm radio}\propto L_{\rm X}^{0.6}$ and $L_{\rm disc}\propto \dot{m}^{2-3}$. The other is the ``efficient'' branch, where $L_{\rm radio}\propto L_{\rm X}^{1.4}$ and $L_{\rm disc}\propto \dot{m}$.

\begin{figure}[!t]
\begin{center}
\includegraphics[angle=270,scale=0.4]{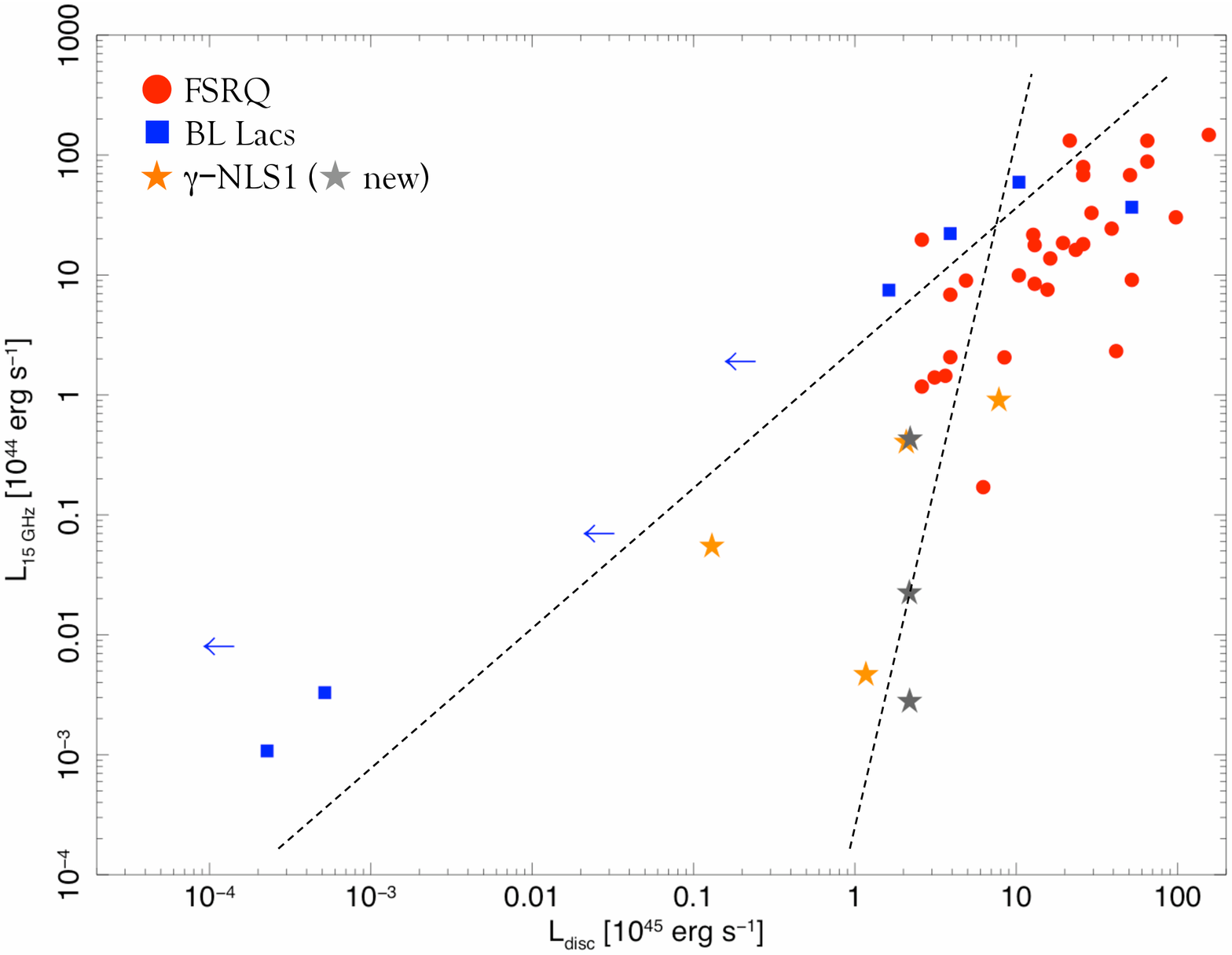} 
\caption{Jet power as from radio measurements vs accretion luminosity. Two dashed lines have the same slopes of the inefficient and efficient branches of Galactic binaries as found by Coriat et al. (2011) are displayed for comparison. See the text for more details.}
\label{fig:xrb}
\end{center}
\end{figure}

The data available for this study does not allow to perform detailed correlations with significant tests, however the Fig.~\ref{fig:xrb} displays some striking similarities with Fig.~5 of Coriat et al. (2011). BL Lac objects seems to sample the inefficient branch, while FSRQ and $\gamma$-NLS1 are on the efficient one. 

Please note that to perform a proper comparison, it is necessary to take into account that the accretion disc in AGN has its peak emission in the ultraviolet band, while in binaries it emits mostly in the X-ray band. 

The difference of disc luminosities between Galactic binaries and AGN is about 9 orders of magnitudes, which can be explained with the mass difference: a few solar masses for binaries and $10^{8-9}M_{\odot}$ for most of the AGN of this sample. On the other hand, the difference of the radio power is about 14 orders of magnitudes, which cannot be explained with the difference in mass. 

Stellar mass black holes are on the similar track of BL Lacs and FSRQs. Interestingly, the $\gamma$-NLS1s, often compared to Galactic black holes in high soft state, occupy a region similar to that of neutron stars in the diagram by Coriat et al. (2011). So, as neutron stars are the low mass sources in the realm of Galactic compact objects, the $\gamma$-NLS1s are the low mass part of the AGN kingdom. See Foschini (2011c) for more details on the similarities between jets in AGN and Galactic binaries.
 
\section{Final remarks}
The discovery of high-energy $\gamma$ rays from NLS1s has perturbed the traditional scenario of AGN with powerful relativistic jets. The new informations carried by adding this class of sources helped to improve our knowledge of the jet mechanisms, although not yet in a definitive way. It is confirmed the existence of two main branches for the AGN with powerful relativistic jets in the framework of the Blandford-Znajek theory: one driven by the accretion and another where the accretion contribution is saturated and is therefore scaled by the mass of the central object. However, when high accretion rate saturates the BZ power, the rise of hydromagnetic wind contributes to increase the jet power. Observations are strongly needed to improve the sample of $\gamma$-NLS1s. 

Interestingly, there is a now striking similarity with similar diagram (accretion vs jet power) in Galactic compact objects. While BL Lac Objects and FSRQs have their like counterpart in Galactic black holes in different states, the $\gamma$-NLS1s now occupy a region similar to that of neutron stars, thus completing the similarity between extragalactic and Galactic classes of compact objects. Although, the difference of 14 orders of magnitudes in the jet power cannot be explained simply with a mass scaling. What is missing?

\begin{acknowledgements}
I would like to thanks L. Maraschi, G. Ghisellini and F. Tavecchio for fruitful discussions. 
This research has made use of the NASA/IPAC Extragalactic Database (NED) which is operated by the Jet Propulsion Laboratory, California Institute of Technology, under contract with the National Aeronautics and Space Administration. This research has made use of data from the MOJAVE database that is maintained by the MOJAVE team (Lister et al., 2009, AJ, 137, 3718). 
\end{acknowledgements}

\label{lastpage}


\begin{thebibliography}{}
\bibitem[Abdo et al. (2009a)]{FERMI} Abdo A.~A. et al., 2009a, ApJ, 707, L142

\bibitem[Abdo et al. (2009b)]{FERMI} Abdo A.~A. et al., 2009b, ApJ, 697, 934

\bibitem[Belloni (2010)]{BELLONI} Belloni T. (ed), 2010, The jet paradigm, Lect. Notes Phys. 794, Springer-Verlag, Berlin. 

\bibitem[Blandford (1976)]{BZNEWTON} Blandford R.~D., 1976, MNRAS, 176, 465

\bibitem[Blandford \& Znajek (1977)]{BZ} Blandford R.~D. \& Znajek R.~L., 1977, MNRAS, 179, 433

\bibitem[Blandford \& Rees (1978)]{BLANDFORD1} Blandford R.~D. \& Rees M.~J., 1978, in: Proceedings of the Pittsburgh Conference on BL Lac Objects, Pittsburgh, April 24-26, 1978, University of Pittsburgh, p. 328

\bibitem[Blandford \& K\"onigl (1979)]{BLANDFORD2} Blandford R.~D. \& K\"onigl A., 1979, ApJ, 232, 34

\bibitem[Blandford \& Payne (1982)]{BP} Blandford R.~D. \& Payne D.~G., 1982, MNRAS, 199, 883

\bibitem[Bloom (2008)]{BLOOM} Bloom S.~D., 2008, AJ, 136, 1533

\bibitem[B\"ottcher \& Dermer (2002)]{BOETTCHER} B\"ottcher M. \& Dermer C.~D., 2002, ApJ, 564, 86

\bibitem[Broderick \& Fender (2011)]{BRODERICK} Broderick J.~W. \& Fender R.~P., 2011, \texttt{arXiv:1105.3769}

\bibitem[Cavaliere \& D'Elia (2002)]{CAVALIERE} Cavaliere A. \& D'Elia V., 2002, ApJ, 571, 226

\bibitem[Coriat et al. (2011)]{CORIAT} Coriat M. et al., 2011, MNRAS, 414, 677

\bibitem[Feigelson \& Nelson (1985)]{ASURV1} Feigelson E.~D. \& Nelson P.~I., 1985, ApJ, 293, 192

\bibitem[Foschini (2011a)]{FOSCHINI1} Foschini L., 2011a, in: Proceedings of the workshop ``Narrow-Line Seyfert 1 Galaxies and Their Place in the Universe'', Milano, April 4-6, 2011, PoS(NLS1) 024 [\texttt{arXiv:1105.0772}].

\bibitem[Foschini (2011b)]{FOSCHINI2} Foschini L., 2011b, \texttt{arXiv:1103.2008}

\bibitem[Foschini (2011c)]{FOSCHINI3} Foschini L., 2011c, \texttt{arXiv:1107.2785}

%\bibitem[Frank et al. (2002)]{FRANK} Frank J. et al., 2002, Accretion power in astrophysics. Cambridge University Press, Cambridge. 

\bibitem[Garofalo et al. (2010)]{GAROFALO} Garofalo D. et al., 2010, MNRAS, 406, 975

\bibitem[Ghirlanda et al. (2011)]{GHIRLANDA} Ghirlanda G. et al., 2011, MNRAS, 413, 852

\bibitem[Ghisellini \& Tavecchio (2008)]{GG3} Ghisellini G. \& Tavecchio F., 2008, MNRAS, 387, 1669

\bibitem[Ghisellini \& Tavecchio (2009)]{GG1} Ghisellini G. \& Tavecchio F., 2009, MNRAS, 397, 985

\bibitem[Ghisellini et al. (2010)]{GG2} Ghisellini G. et al., 2010, MNRAS, 402, 497

\bibitem[Ghisellini et al. (2011)]{GG4} Ghisellini G. et al., 2011, MNRAS, 414, 2674

\bibitem[Ghosh \& Abramowicz (1997)]{GHOSH} Ghosh P. \& Abramowicz M.~A., 1997, MNRAS, 292, 887

\bibitem[Gliozzi et al. (2010)]{GLIOZZI} Gliozzi M. et al., 2010, ApJ, 717, 1243

\bibitem[Goldreich \& Julian (1969)]{GOLDREICH} Goldreich P. \& Julian W.~H., 1969, ApJ, 157, 869

\bibitem[Gu \& Chen (2010)]{GU} Gu M. \& Chen Y., 2010, AJ, 139, 2612

\bibitem[Isobe et al. (1986)]{ASURV2} Isobe T. et al., 1986, ApJ, 306, 490

\bibitem[Komatsu et al. (2011)]{KOMATSU} Komatsu E. et al., 2011, ApJS, 192, 18

\bibitem[K\"ording et al. (2006)]{KOERDING} K\"ording et al., 2006, MNRAS, 369, 1451

\bibitem[Laor (2000)]{LAOR} Laor A., 2000, ApJ, 543, L111

\bibitem[Lavalley et al. (1992)]{ASURV3} Lavalley M.~P. et al., 1992, Bull. Am. Astron. Soc. 24, 839

\bibitem[Lovelace (1976)]{LOVELACE} Lovelace R.~V.~E., 1976, Nature, 262, 649 

\bibitem[Lister et al. (2009)]{LISTER} Lister M.~L. et al., 2009, AJ, 137, 3718

\bibitem[Macdonald \& Thorne (1982)]{BZMT} Macdonald D. \& Thorne K.~S., 1982, MNRAS, 198, 354

\bibitem[Maraschi \& Tavecchio (2003)]{MARASCHI} Maraschi L. \& Tavecchio F., 2003, ApJ, 593, 667

\bibitem[Meier (1999)]{MEIER} Meier D.~L., 1999, ApJ, 522, 753

\bibitem[Moderski \& Sikora (1996)]{MODERSKI} Moderski R. \& Sikora M., 1996, MNRAS, 283, 854

\bibitem[Osterbrock \& Pogge (1985)]{OSTERBROCK} Osterbrock D.~E. \& Pogge R.~W., 1985, ApJ, 297, 166

\bibitem[Ozernoy \& Usov (1973)]{OZERNOY} Ozernoy L.~M. \& Usov V.~V., 1973, Ap\&SS, 25, 149

\bibitem[Penrose (1969)]{PENROSE} Penrose R., 1969, Rivista Nuovo Cimento, 1, 257

\bibitem[Phinney (1983)]{PHINNEY} Phinney E.~S., 1983, in: Astrophysical Jets: Proceedings of the International Workshop, eds A. Ferrari \& A.~G. Pacholczyck, NATO--ASI Series vol. 103, D. Reidel Publishing Co., Dordrecht, p. 201.

\bibitem[Piddington (1970)]{PIDDINGTON} Piddington J.~H., 1970, MNRAS, 148, 131

\bibitem[Punsly \& Coroniti (1990)]{PUNSLY} Punsly B. \& Coroniti F.~V., 1990, ApJ, 354, 583

\bibitem[Ruffini \& Wilson (1975)]{RUFFINI} Ruffini R. \& Wilson J.~R., 1975, Physical Review D, 12, 2959

\bibitem[Sikora et al. (2007)]{SIKORA} Sikora M. et al., 2007, ApJ, 658, 815

\bibitem[Stocke et al. (2011)]{STOCKE} Stocke J.~T. et al., 2011, ApJ, 732, 113

\bibitem[Sturrock \& Barnes (1972)]{STURROCK} Sturrock P.~A. \& Barnes C., 1972, ApJ, 176, 31

\bibitem[Thorne et al. (1986)]{THORNE} Thorne K.~S., Price R.~H., Macdonald D.~A. (eds), 1986, Black holes: The membrane paradigm, Yale University Press, New Haven.

\bibitem[Tombesi et al. (2010)]{TOMBESI} Tombesi F. et al., 2010, ApJ, 719, 700

\bibitem[Woo \& Urry (2002)]{WOOURRY} Woo J.-H. \& Urry C.~M., 2002, ApJ, 581, L5 (Erratum: ApJ, 583, L47)

\end{thebibliography}
\end{document}